\documentclass[12pt]{iopart}
\usepackage{graphicx, iopams}

\expandafter\let\csname equation*\endcsname\relax

\expandafter\let\csname endequation*\endcsname\relax

\usepackage{amsmath}

\begin{document}

\title{Bohmian trajectories in an entangled two-qubit system}

\author{A.C. Tzemos, G. Contopoulos and C. Efthymiopoulos}

\address{Research Center for Astronomy and 
Applied Mathematics of the Academy of 
Athens - Soranou Efessiou 4, GR-11527 Athens, Greece}
\ead{thanasistzemos@gmail.com,
gcontop@academyofathens.gr, 
cefthim@academyofathens.gr}
\vspace{10pt}

\begin{abstract}
In this paper we examine the evolution of Bohmian 
trajectories in the presence of quantum entanglement.
We study a simple two-qubit system composed 
of two coherent states 
 and investigate the impact of quantum 
entanglement on chaotic and ordered trajectories via 
both numerical and analytical calculations.
\end{abstract}

\submitto{\JPA}

\section{Introduction}
Bohmian Quantum Mechanics (BQM) is one of the main alternative 
interpretations of Quantum Mechanics (QM) (\cite{Bohm, BohmII}), 
where quantum particles follow certain trajectories
in spacetime, 
in sharp contrast with Standard Quantum Mechanics (SQM),
where the notion of particle trajectory does not exist. 
However they both predict the same experimental results. 
In BQM  the usual Schr\"{o}dinger's equation (SE) still
governs the evolution  of the wave function $\Psi$
which in turn guides the particle positions via 
a set of nonlinear  first order in time 
equations of motion,  the Bohmian equations.

Quantum entanglement (QE) is the basic property that makes 
quantum systems behave differently from classical systems
\cite{horodecki2009quantum}, \cite{mintert2005measures}.
QE plays a key role in Quantum Information 
and Computation Theory, since it is useful in many applications such as computing algorithms, quantum teleportation schemes and public key distribution protocols \cite{nielsen2004quantum}.
From a theoretical perspective QE is the
manifestation of
the non local nature of QM. Consequently 
it has a central role in Bohmian Mechanics, where it 
has been studied in different 
frameworks, such as the theoretical and 
experimental study of the relation between 
Bohmian trajectories and  quantum measurements 
\cite{durt2002bohm, braverman2013proposal,norsen2014weak, mahler2016experimental},
the dynamics of interacting many body  systems 
\cite{elsayed2018entangled}, the relation 
between chaos and entanglement  in the
case of stationary states (focusing on three-partite systems) \cite{cesa2016chaotic}, the dynamics of dissipative bipartite systems \cite{de2012bohmian}
and the study of correlations between 
spin-1/2 quantum rotors in comparison with 
SQM \cite{ramvsak2012spin}.

In this paper we investigate the effect of QE on the Bohmian 
trajectories of  a simple system composed of coherent 
states of two independent harmonic oscillators.
Namely we study the Bohmian trajectories on the $x-y$ plane 
of a composite system, whose subsystems evolve in the $x$ 
and $y$ coordinates respectively. Our system is convenient 
for the numerical study of  Bohmian trajectories. Furthermore 
it is characterized by complex dynamics and exhibits different 
behavior for different values of the physical parameters. 
Finally, it gives us the opportunity 
to work with analytical relations for
the entanglement. 

Here we focus on how QE influences the behaviour of the quantum trajectories, 
in the case of simple bi-partite quantum systems, whose entanglement is well understood
and unambiguously quantified.
This is the first step of a research plan aiming to define indicators and measures of quantum entanglement based
  on the Bohmian trajectories. The construction of such 
quantities should be helpful for the study of multipartite entangled 
systems, where the quantification of entanglement remains an open problem. In fact, the  
Bohmian trajectories allow us to transform the question of how to measure entanglement to a measurement of the level of the coupling between the variables in the Bohmian equations
of motion.

We find that in our system the basic criterion for the 
behaviour of Bohmian trajectories is the ratio of the angular 
frequencies $\omega_1/\omega_2$. When this ratio is irrational 
we observe chaotic trajectories, while when it is 
rational we observe periodic trajectories. 
In the case of incommensurable frequencies 
we find that entanglement 
is necessary  for the emergence of chaos.
In the case of commensurable frequencies
we find that the motion is always periodic
even if for small intervals of time can be
described as effectively chaotic. 
Finally, in the case of the isotropic oscillators 
we see that the increase of
entanglement confines the range of the periodic motion and
changes its Fourier spectrum.

The present paper is organized in the following
way: In Section. \ref{ch} we present the  system 
of two entangled qubits.
Then we  compute some standard measures allowing to quantify 
entanglement in our system. Section \ref{Bohmeq} discusses 
the Bohmian equations of motion that govern our 
system with a reference to the main mechanism
responsible for the production of chaos in 
2-d Bohmian trajectories. Section \ref{traj} deals with the effect
of entanglement on the evolution of Bohmian trajectories,
firstly in the case of incommensurable frequencies
and then  of commensurable frequencies.
Finally we study the extreme case of the isotropic
oscillators and discuss its unique features.
 In Section \ref{conc} we summarize our
results and conclusions.

\section{Two state system and entanglement} \label{ch}
\subsection{Hamiltonian and Coherent States} 
We consider a system of two uncoupled particles, 
of masses $m_x$ and $m_y$ moving in coordinates 
$x, y$ under the influence of an external harmonic
potential with frequencies $\omega_x$ 
and $\omega_y$. The system is described by the Hamiltonian:
\begin{eqnarray}
H=\frac{p_x^2}{2m_x}+\frac{p_y^2}{2m_y}+\frac{1}{2}m_x\omega_x^2x^2+\frac{1}{2}m_y
\omega_y^2y^2
\end{eqnarray}
We  examine states formed by combinations of coherent states for the two 
particles. Coherent states are defined as the
eigenstates of the annihilation operator $\hat{\alpha}$ 
associated to the eigenvalue
$A$:

\begin{equation}
\hat{\alpha}|\alpha(t)\rangle=A(t)|\alpha(t)\rangle,
\end{equation}
where $A(t)=|A(t)|\exp(i\phi(t))$.
The wavefunction of a coherent state in the position
representation has the form:
\begin{equation}
Y(x,t; A_0,\sigma,\omega, m)=\Bigg(\frac{m\omega}{\pi\hbar}\Bigg)^{\frac{1}{4}}
\exp\Bigg[-\frac{m\omega}{2\hbar}\Bigg(x-\sqrt{\frac{2\hbar}{m\omega}}
\Re[A(t)]\Bigg)^2+i\Bigg(\sqrt{\frac{2m\omega}{\hbar}}
\Im[A(t)]x+\xi(t)\Bigg)\Bigg],
\end{equation}\label{cs}
where
\begin{eqnarray}
\xi(t)=\frac{1}{2}\Big[|A_0|^2
\sin(2(\omega t-\sigma))-\omega t\Big]
\\
\Re[A(t)]=|A_0|\cos(\sigma-\omega t)\\
\Im[A(t)]=|A_0|\sin(\sigma-\omega t)
\end{eqnarray}
with $\sigma=\phi(0)$ the initial phase of $A$ and $A_0=A(0)$
since in Schr\"{o}dinger's picture
we have:
\begin{eqnarray}
A(t)&=\exp(-i\omega t)A_0\\&=
\exp\Big[-i\big(\omega t-\phi(0)\big)\Big]|A_0|.
\end{eqnarray}
Two arbitrary coherent states 
$\alpha_1, \alpha_2$ are in general not orthogonal
to each other since
\begin{equation}
|\langle\alpha_2|\alpha_1\rangle|^2=
\exp(-|A_1-A_2|^2).
\end{equation}
However, the overlapping decreases exponentially with their
distance in the phase space
(see\cite{garrison2008quantum}).
In our numerical experiments we choose the difference $|A_1-A_2|$ to be large. This  creates effectively  a two `qubit' system in the position representation.

\subsection{Entangled Qubits}

\begin{figure}
\centering
\includegraphics[scale=0.35]{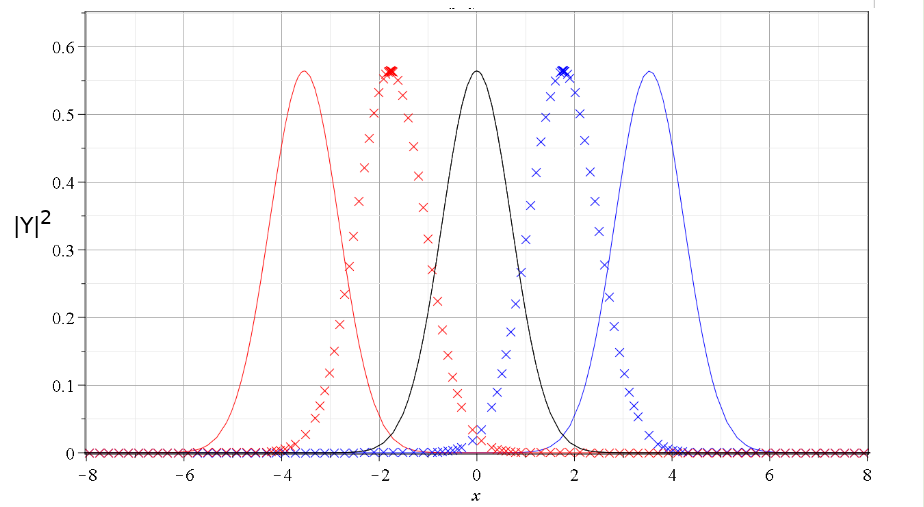}
\caption{The basis states of the qubit.
At $t=0$ we have the solid blue curve on the 
right hand and the solid red curve on the 
left hand. They represent the absolute square of the wavefunctions of the coherent states $|R\rangle$
and $|L\rangle$ correspondingly. 
At $t=\pi/3$ the curves move
towards $x=0$ (diagonal  cross blue and red
curves), while at $t=\pi$ they become identical 
at the center of the oscillation $x=0$. The 
oscillation period here is $T=2\pi$. 
(For both curves $|A_0|=5/2, \omega_x=1$,
while  $\sigma_x=0$ for the blue one
and $\sigma_x=\pi$ for the red one.}
\label{basis_states}
\end{figure}

We work with  quantum states described in the position representation by wavefunctions of the form:

\begin{eqnarray}
\Psi(x,y,t)=c_1Y_R(x,t)Y_L(y,t)+
c_2Y_L(x,t)Y_R(y,t)\label{psiplus}
\end{eqnarray} or
\begin{eqnarray}
\Phi(x,y,t)=c_1Y_R(x,t)Y_R(y,t)+
c_2Y_L(x,t)Y_L(y,t)\label{phiplus}
\end{eqnarray}
where $|c_1|^2+|c_2|^2=1$ and
\begin{eqnarray}
Y_R(x,t)\equiv Y(x,t;\omega=\omega_x,m=m_x,\sigma=\sigma_x)\\
Y_R(y,t)\equiv Y(y,t;\omega=\omega_y,m=m_y,\sigma=\sigma_y)\\
Y_L(x,t)\equiv Y(x,t;\omega=\omega_x,m=m_x,\sigma=\sigma_x+\pi)\\
Y_L(y,t)\equiv Y(y,t;\omega=\omega_y,m=m_y,\sigma=\sigma_y+\pi)
\end{eqnarray}
Setting  $\sigma_x=0$ and $\sigma_y=0$, 
the symbols R and L refer to the right or 
left position of the Gaussian wavepacket
of a one-dimensional coherent state in $x$
or $y$ direction, with respect to the center
of the oscillation at time $t=0$. 
The initially right or left position in physical space defines the
basis states $\{|R\rangle, |L\rangle\}$ of a qubit
\begin{equation}
|Q\rangle=a|R\rangle+b|L\rangle
\quad |a|^2+|b|^2=1,
\end{equation}
under the assumption that
their overlap in phase space is sufficiently small. An example is
given in Fig.~\ref{basis_states}, 
corresponding to $|A_x(0)|=|A_y(0)|=|A_0|=5/2$. This gives a negligible overlap, since
$|\langle L|R\rangle|\,\sim
\mathcal{O} (10^{-6})$.

\subsection{Entanglement}\label{ent}
The Von Neumann entropy ($VNE$) is the quantum
mechanical extension of Gibbs entropy. For 
a state described by the density matrix
$\rho$ (in a given basis) we have:
\begin{equation}
VNE=-tr[\rho\ln(\rho)]
\end{equation}
which, for given eigenvectors ${|\Lambda_i\rangle}$
and eigenvalues ${\lambda_i}$ of $\rho$, reduces to 
\begin{equation}
VNE=-\sum_i \lambda_i\ln(\lambda_i).
\end{equation}
A non vanishing $VNE$ reflects the departure 
of a system from purity, since for a pure
state we have $VNE=0$ ($\rho$ has one
eigenvalue equal to 1). On the other hand
$VNE$ takes a maximal value equal to $\ln(N)$ 
for $N$-dimensional Hilbert space in the
case of  a maximally mixed state. We note 
that the $VNE$ is preserved under unitary
evolution since
$VNE(\rho)=VNE(U \rho U^\dagger)$, for any
a unitary transformation $U$.

 The $VNE$ can be used for the quantification
of QE. In particular we can apply a partial
trace operation over the degrees of freedom 
of one subsystem and then calculate the $VNE$
of the reduced density matrix which describes 
the remaining subsystem. The $VNE$ entropy of 
the reduced density matrix is called entanglement 
entropy ($EE$), and is a reliable measure of 
bipartite entanglement
\begin{equation}
EE\equiv VNE_a=VNE_b=-tr[\rho_A\ln(\rho_A)]
=-tr[\rho_B\ln(\rho_B)].
\end{equation}

The calculation of $VNE$ is in general
a demanding task, since one needs to
diagonalize the density matrix, whose
dimension is in principle very large.
This means that in general an analytical calculation
of $EE$ is  difficult (see for example\cite{makarov2018coupled}).

For a generic density matrix $\rho$ the linear entropy $LE$ is defined as 
\begin{equation}
LE=1-\tr(\rho^2).
\end{equation}
For a given $N\times N$ density matrix, the 
values of the $LE$ lie between $0$ and $1-1/N$ 
for a pure state and a maximally mixed state 
respectively.
In fact $LE$ is an approximation of the $VNE$ 
and its calculation is simpler
than that of $VNE$, since it does not require 
diagonalization of the density matrix.
Consequently in the case of a pure state 
\begin{equation}
\Psi(x,y)=\sqrt{R(x,y)}e^{iw(x,y)}
\end{equation}  we can use as a measure 
of the bipartite entanglement the $LE$ of 
the reduced density matrix:
\begin{eqnarray}
\nonumber LE_{RED} &=1-\tr(\rho_A^2)
=1-\tr(\rho_B^2)\\&
=1-\int dxdydx'dy'\Psi(x,x')
\Psi^{\dagger}(y,x')\Psi(x,y')
\Psi^{\dagger}(y,y').
\end{eqnarray}

\begin{figure}[hb]
\centering
\includegraphics[scale=0.28]{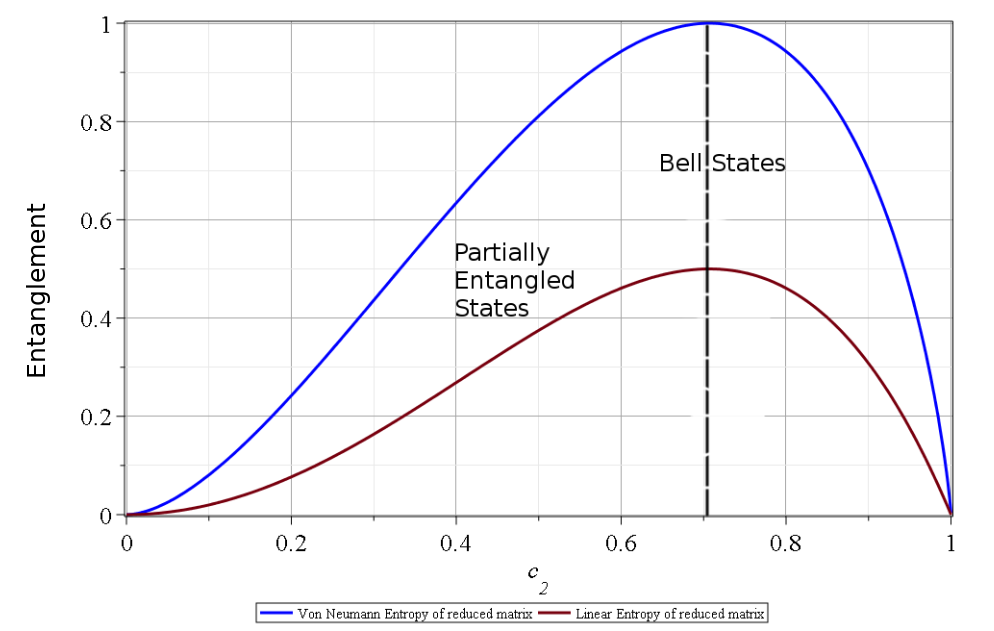}
\caption{
The entanglement as a function of $c_2$, 
measured by the $EE$ of Eq.~(\ref{EE}) (blue curve) and  
$LE_{RED}$ of Eq.~(\ref{LEQ}) (bordeuax curve). It is common for both states $\Psi$ and $\Phi$.
We observe that $EE=0$ for $c_2=0$ and $c_2=1$ , 
something expected since in that cases we 
have product states. The maximum value of
$LE_{RED}^{max}=1/2$ corresponds to $c_2=\frac{\sqrt{2}}{2}$,
namely in the case of a Bell state
(maximally entangled state). In that case
the $EE_{max}$  is equal to $\ln(2)$, or
equal to 1 if we use 2 as the basis of
the logarithm (as we have done 
here), which is the case in quantum information
theory. All of the 
other values of $c_2$ refer to partially 
entangled states. }
\label{c2ande}
\end{figure}

In order to study entanglement in Bohmian systems, 
Zander and Plastino showed in 
\cite{zander2018revisiting} that
 $LE_{RED}$ can be written as the sum 
of two quantities:
\begin{equation}
LE_{RED}=LE_c+LE_p,
\end{equation}
where 
\begin{equation}\label{ec}
LE_c=1-\int dxdydx'dy'[R(x,x')
R(y,x')R(x,y')R(y,y')]^{\frac{1}{2}}
\end{equation}
is the `configuration entanglement' and
\begin{eqnarray}\label{ep}
\nonumber \fl LE_p=\int dxdydx'dy'
&\{1-\exp[i\Big(w(x,x')
-w(y,x')-w(x,y')-w(y,y'\Big)]\}\\&
\times
[R(x,x')
R(y,x')R(x,y')R(y,y')]^{\frac{1}{2}}
\end{eqnarray}
is the `phase entanglement'.
The configuration entanglement expresses the 
lack of factorizability of the probability
density  $R(x,y)$, namely $
R(x,y)\neq R_1(x)R_2(y)$, while
the phase entanglement expresses the lack
of additivity of the phase $w(x,y)$:
\begin{equation}
w(x,y)\neq w_1(x)+w_2(y).
\end{equation}
In general, the analytical calculation of (\ref{ec})
and (\ref{ep}) is  difficult 
and one needs to proceed numerically with algorithms
like Cuhre or Monte-Carlo.
However, in our case we study a system of
two non-interacting subsystems. Consequently
its entanglement remains constant over time.
This means that a proper choice of the value
of time $t$ (for a given wavefunction) can 
simplify the calculations. In our case if 
we assume that $c_1, c_2$ are real, we can compute 
the entanglement at $t=0$, 
where the imaginary part of the wavefunctions
$\Phi$ and $\Psi$
is equal to $0$ and consequently $LE_{RED}=LE_c$. In
the case of $\Phi$ the multiple 
integral can be solved numerically. In the case of 
$\Psi$ we managed
to find the solution analytically:
\begin{eqnarray}
&\fl LE_c^{\Psi}=1-\Big[{c_{1}}^{4}-{c_{2
}}^{4}+\left( -4{c_{1}}^{3}c_{2}+4
{c_{1}}^{2}{c_{2}}^{2}-4c_
{1}{c_{2}}^{3} \right) {{\rm e}^{-4{\alpha_{0}}^{2}}}
+2{c_{1}}^{2}{c_{2}}^{2}{{\rm e}^{-8{\alpha_{0}}^{2}}})\Big],
\end{eqnarray}
where $a_0\equiv|A_0|$. We note the fact
that for all $\alpha_0$ $LE_{RED}=LE_c$ is 
independent from $\omega_x, \omega_y$.
In our case (large $\alpha_0$ and 
$c_1, c_2\in[0,1]$) with $c_1^2+c_2^2=1$
we find
\begin{equation}
LE_c=1- \left(  {c_{1}}^{4}+{c_{2
}}^{4})=2c_2^2(1-c_2^2
\right)\label{LEQ}
\end{equation}
Note that, for large $a_0$ our system becomes equivalent to a two-qubit spin system described by a $4\times 4$ density 
matrix in the standard 
basis $(\{|0\rangle,|1\rangle\})$, if we 
correspond $|0\rangle\to|R\rangle$ and 
$|1\rangle\to|L\rangle$. Then Eq.~(\ref{LEQ})
is nothing else than the linear entropy 
of its reduced density matrix. However 
in this case it is trivial to calculate 
the $VNE$ of the reduced density matrix for both states  (\ref{psiplus}) and (\ref{phiplus}), the so called entanglement entropy $EE$:
\begin{equation}\label{EE}
EE=-\Big(|c_1|^2\ln(|c_1|^2)+|c_2|^2\ln(|c_2|^2))
=(1-|c_2|^2)\ln(1-|c_2|^2)+|c_2|^2\ln(|c_2|^2))
\end{equation}
Figure~\ref{c2ande} shows
the $LE_{RED}$ (Eq.~(\ref{LEQ})) and $EE$ (Eq.~(\ref{EE}))  for different values of 
the coefficient $c_2$. Both of them exhibit the same behaviour
for all values of $c_2$, hence $LE_{RED}$ is a reliable
measure for the quantification of entanglement.
In particular, the maximum entanglement takes 
place when $c_2=\sqrt{2}/2$ (Bell states).

\section{Bohmian equations and nodal points}\label{Bohmeq}
The Bohmian trajectories guided by
a wavefunction $\Psi=\Psi_R+i\Psi_I$ are found via the
Bohmian equations
\begin{eqnarray}\label{bohmeq}
m_i\frac{dx_i}{dt}=\frac{\hbar}{G}\Big(\frac{\partial\Psi_I}
{\partial x_i}\Psi_R-\frac{\partial \Psi_R}
{\partial x_i}\Psi_I\Big),\quad i=1,2,\dots
\end{eqnarray}
with $G=\Psi_R^2+\Psi_I^2$. Hereafter we work with $\hbar=m_1=m_2=1$.
The Bohmian equations for the state $\Psi$ are:

\begin{eqnarray}
&\fl\frac{dx}{dt} =-\frac {\sqrt {2\omega_x}a_{0}
 \left[ A\cos \left(\omega_{x}t \right) +B \sin \left( \omega_{x}t
 \right)  \right] }{G}\label{rational1}\\&
\fl\frac{dy}{dt}=\frac {\sqrt {2\omega_y}a_{0}
 \left[ A 
\cos \left( \omega_{y}t \right) +B\sin\left( \omega_{y}t \right) \right] }{G}\label{rational2}
\end{eqnarray}
and for the state $\Phi$ are:
\begin{eqnarray}
&\fl\frac{dx}{dt} =-\frac {\sqrt {2\omega_x}a_{0}
 \left[ C\cos \left(\omega_{x}t \right) +D \sin \left( \omega_{x}t
 \right)  \right] }{G'}\\&
\fl\frac{dy}{dt}=-\frac {\sqrt {2\omega_y}a_{0}
 \left[ C 
\cos \left( \omega_{y}t \right) +D\sin\left( \omega_{y}t \right) \right] }{G'}
\end{eqnarray}
where
\begin{eqnarray}
A= 2c_{1}c_{2}{{\rm e}^{2
f_{x}+2f_{y}}}\sin \left( 2(g_{x}-g_{y})\right)\\
B=c_1^2e^{4f_x}-c_2^2e^{4f_y}\\
G=2c_{1}c_{2}{{\rm e}^{2f_{x}+2f_{y}}
}\cos \left( 2(g_{x}-g_{y}) \right) +{{\rm e}^{4f_{y}}}{c_{2}}^{2
}+{{\rm e}^{4f_{x}}}{c_{1}}^{2}\\
C= 2c_{1}c_{2}{{\rm e}^{2
f_{x}+2f_{y}}}\sin \left( 2(g_{x}+g_{y})\right)\\
D={c_{1}}^{2}{{\rm e}^{4f_{x}+4f_{y}}}-{c_{2}}^{2}\\
G'={c_{1}}^{2}{{\rm e}^{4f_{x}+4f_{y}}}+2c
_{1}c_{2}{{\rm e}^{2f_{x}+2f_{y}}}\cos \left( 2(g_{x}+g_{y
}) \right) +{c_{2}}^{2}
\end{eqnarray}
with
\begin{eqnarray}
f_{x}=\sqrt {2\omega_{x}} a_{0}\cos \left( \omega_{x}
\,t \right)x,\quad f_{y}=\sqrt {2\omega_{y}}a_{
0}\cos \left( \omega_{y}\,t \right) y,\\
g_{x}=\sqrt {2\omega_{x}}a_{0}\,\sin \left( \omega_{x}\,t
\right) x,\quad g_{y}=\sqrt {2\omega_{y}}a_{0}\,\sin \left( \omega_{y}\,t
\right) y
\end{eqnarray}

A useful quantity for the identification of chaos
is the ``finite time Lyapunov characteristic number''
LCN.  
If $\xi_k$ is the length of 
the deviation vector between two nearby 
trajectories at the time 
$t = \kappa t_0,\, \kappa = 1, 2,\dots$,
then the quantity   
\begin{equation}
\alpha_\kappa=\ln\Big(\frac{\xi_{\kappa+1}}
{\xi_\kappa}\Big)\label{str}
\end{equation}
is the so called ``stretching number''
and the ``finite time Lyapunov characteristic number" 
is given by the equation:
\begin{equation}
\chi=\frac{1}{\kappa t_0}\sum_{i=1}^
\kappa\alpha_i\label{flcn}
\end{equation}
The LCN is the limit of $\chi$ when $\kappa\to\infty$ 
and the ratio $\frac{\xi_{\kappa+1}}
{\xi_\kappa}$ is computed by the variational equations 
of motion (see \cite{voglis1994invariant, Contopoulos200210}).
LCN is positive for chaotic trajectories
and equal to zero for ordered trajectories.

In  previous works we made a detailed investigation 
of order and chaos in BQM
\cite{efthymiopoulos2006chaos, 
PhysRevE.79.036203, tzemos2018integrals, tzemos2018origin}.
In the close neighbourhood of nodal points, which are 
defined as 
solutions of the system:
\begin{eqnarray}
\Re\Psi=\Im\Psi=0,
\end{eqnarray}
the Bohmian particles evolve very fast and 
form spirals around them. In a frame comoving with
a nodal point, the 
nodal point is
accompanied by a second stationary point 
of the flow, the X-point. At the X-point the 
Bohmian velocity
becomes equal to that of the moving node 
\begin{eqnarray}
\frac{dx}{dt}=V_{xnod},\quad \frac{dy}{dt}=V_{ynod}.
\end{eqnarray} Together they 
form the nodal point-X-point 
complex (NPXPC), a characteristic 
geometrical structure of the Bohmian 
flow in the vicinity of
a certain moving nodal point. 
We have shown that the NPXPC
are responsible for the emergence of chaos in two 
and three dimensional Bohmian systems (see also \cite{Wisniacki1, Wisniacki2}). Whenever
a trajectory approaches a NPXPC it gets scattered 
by the X-point and the stretching number 
undergoes a positive shift. The cumulative action
of NPXPCs on the trajectories produces chaos: two arbitrarily close initial conditions produce trajectories whose distance grows exponentially in time. On the other hand, the trajectories that do not interact
with the NPXPCs are ordered.

In our case the nodal points of the $\Psi$ (Eq.~\ref{psiplus})
read:
\begin{eqnarray}
\label{xnod}&x_{nod}={\frac {\sqrt {2}
\left( k\pi\,\cos \left( 
\omega_{y}\,t \right) +\sin \left( 
\omega_{y}\,t \right) \ln  \left( 
\left| {\frac {c_{1}}{c_{2}}} \right|  
\right)  \right) }{4\sqrt {\omega_{x}}a_{0}\,\sin \left(  
\omega_{xy}  t \right) } }\\&
\label{ynod}y_{nod}={\frac {\sqrt {
2} \left(k\pi\, \cos \left( \omega_{x}t 
\right) +\sin \left( \omega_{x}t \right) 
\ln  \left(  \left| 
{\frac {c_{1}}{c_{2}}} \right|  \right)  
\right) }{4\sqrt {\omega_{y}}a_{0}\,\sin 
\left( \omega_{xy}\,t \right) }}
\end{eqnarray}
with $k\in Z $, $k$ even for $c_1c_2<0$
or odd for $c_1c_2>0$ and $\omega_{xy}\equiv \omega_x-\omega_y$. Consequently there 
exist infinitely many nodal points 
in space.
Similarly the nodal points in the case of $\Phi$ (Eq.~\ref{phiplus}) read:

\begin{eqnarray}
& x_{nod}={\frac {\sin \left( \omega_{x}t \right) \sqrt {2}
\left( k\pi\,\cos
\left( \omega_{y}\,t \right)+\ln  \left( \Big|{\frac {  c_{1}  }{  c_{2}
}}\Big| \right) \sin \left( \omega_{y}\,t \right)  \right) }{2\sqrt {
\omega_{x}}a_{0} \left( \cos \left( \omega_{y}t \right) -\cos
\left( t \left( 2\,\omega_{x}-\omega_{y} \right)  
\right)  \right) }}\label{xnod2}\\&
y_{nod}=-{\frac {\sqrt {2}\left( k\pi\cos \left( \omega_{x}
\,t \right) +\sin \left( \omega_{x}\,t \right) \ln  \left( 
\left| {\frac {c_{1}}{c_{2}}} \right|  
\right)  \right)}{4\sqrt {\omega_{y}}a_{0}\sin \left(  \omega
_{xy} t  \right) }  }\label{ynod2}
\end{eqnarray} where again
  $k$ is even for $c_1c_2<0$ or odd for $c_1c_2>0$.
Equations (\ref{xnod}, \ref{ynod},
\ref{xnod2}) and (\ref{ynod2})
show  that in 
general there exist nodal points evolving
 in space-time for different $\omega_x, 
\omega_y$ and various amounts of entanglement 
(different $c_2$). The state $\Psi$ gives:
\begin{equation}
\frac{y_{nod}}{x_{nod}}=\frac{k\pi\, \cos 
\left( \omega_{x}t \right) +\sin \left( \omega_{x}t \right) \ln  \left(  \left| 
{\frac {c_{1}}{c_{2}}} \right|  \right)}{ k\pi\,\cos \left( 
\omega_{y}\,t \right) +\sin \left( \omega_{y}\,t \right) \ln  \left( 
 \left| {\frac {c_{1}}{c_{2}}} \right|  \right)  }
\end{equation}
while the state $\Phi$ gives
\begin{equation}
\frac{y_{nod}}{x_{nod}}=\frac{-  \sqrt {\omega_{x}}
\left( \cos \left( \omega_{x}t \right) k\pi +\sin \left( \omega_{
x} t \right) \ln  \left(  \left| {\frac {c_{1}}{c_{2}}} \right| 
 \right)  \right) \Big( \cos
 \left( \omega_{y}t \right) -\cos \left( (2\omega_{x}\!-\!\omega_{y})
 t \right)  \Big) 
}{2\sqrt {\omega_{y}}\sin \left(  \omega_{xy} t 
\right)  \sin(\omega_xt)\Big[k\pi \cos 
\left( \omega_{y}t \right) +\sin \left( \omega_{y} t
 \right)\! +\!\ln  \left(  \left| {\frac {c_{1}}{c_{2}}} \right|  \right)\Big]  
}
\end{equation}
Finally we note that with the tranfsormation $(t\to -t,k'\to -k)$ one finds the same values of $x_{nod}, y_{nod}$ in both cases. Thus, the same nodal points appear in the plane $(x,y)$ at
the times $t$ and $-t$.

\section{Trajectories}\label{traj}

\subsection{Incommensurable frequencies}
\begin{figure}[h]
\centering
\includegraphics[scale=0.25]{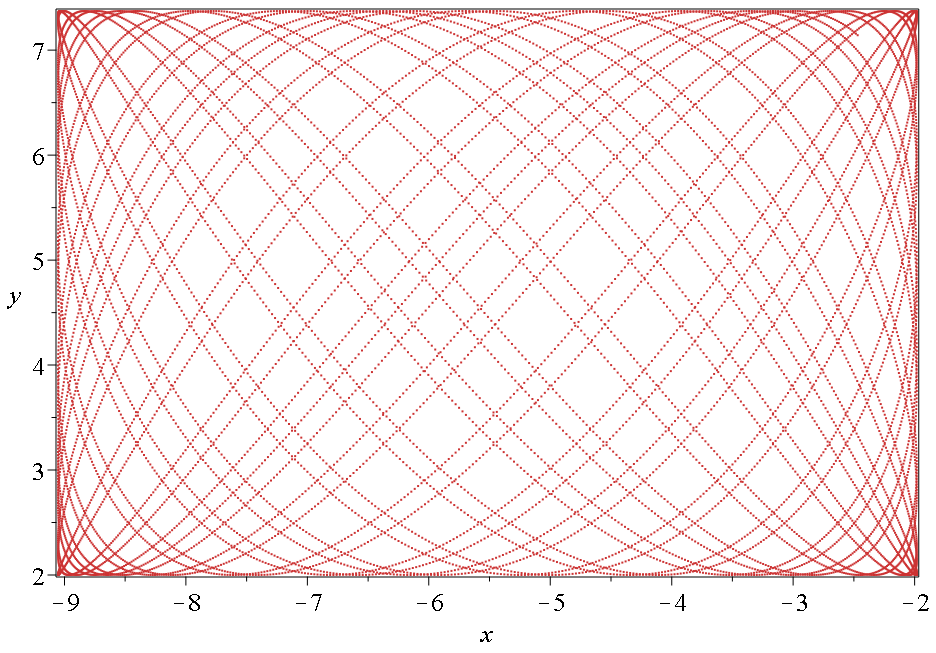}
\caption{A Bohmian trajectory in the 
case of the product state 
$Y_{RL}=Y_RY_L$ for $t\in[0,100]$. 
It is a typical Lissajous curve. $(x(0)=-2, 
y(0)=2, \omega_x=1, \omega_y=\sqrt{3})$}
\label{product_RL}
\end{figure}

When the oscillators have incommensurable 
 frequencies, in the extreme case of a product state the 
trajectories are of Lissajous type. For example  Fig.~\ref{product_RL} shows a
trajectory in the 
state $Y_R(x,t)Y_L(y,t)$ ($c_2=0$). In this case 
the Bohmian equations are decoupled, 
the nodal points are at infinity and 
the motion is ordered.

\begin{figure}[h]
\centering
\includegraphics[scale=0.25]{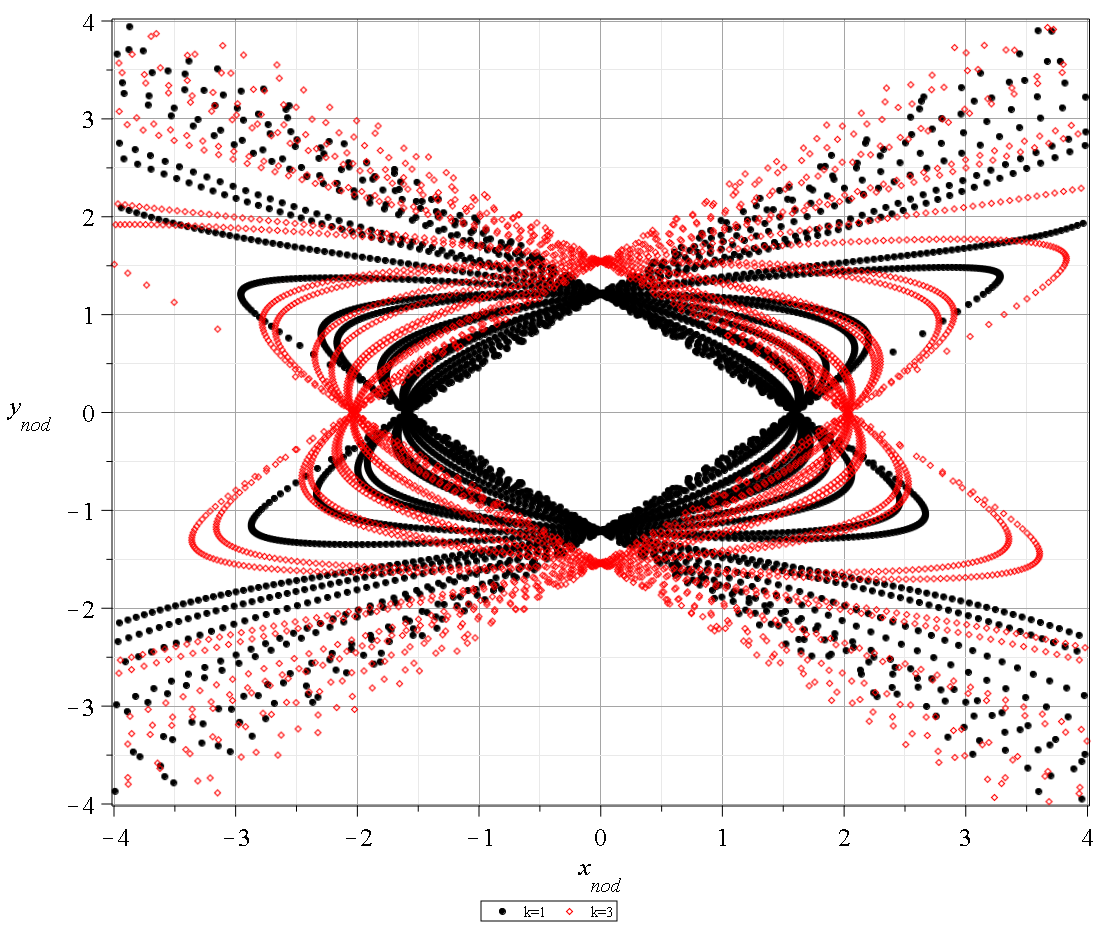}
\caption{
The nodal trajectories of the state $\Psi$ for $k=1$ (black dots) and $k=3$ (red squares)
 for $t\in[0,100]$.
($\omega_x=1,\omega_y=\sqrt{3}, 
c_2=2\times 10^{-5}$). 
Entanglement brings the nodal points
(and consequently the X-points) into
the region where the wavefunction has strong 
support and produces chaos.}
\label{diaforetika_k}
\end{figure}

\begin{figure}[h]
\centering
\includegraphics[scale=0.3]{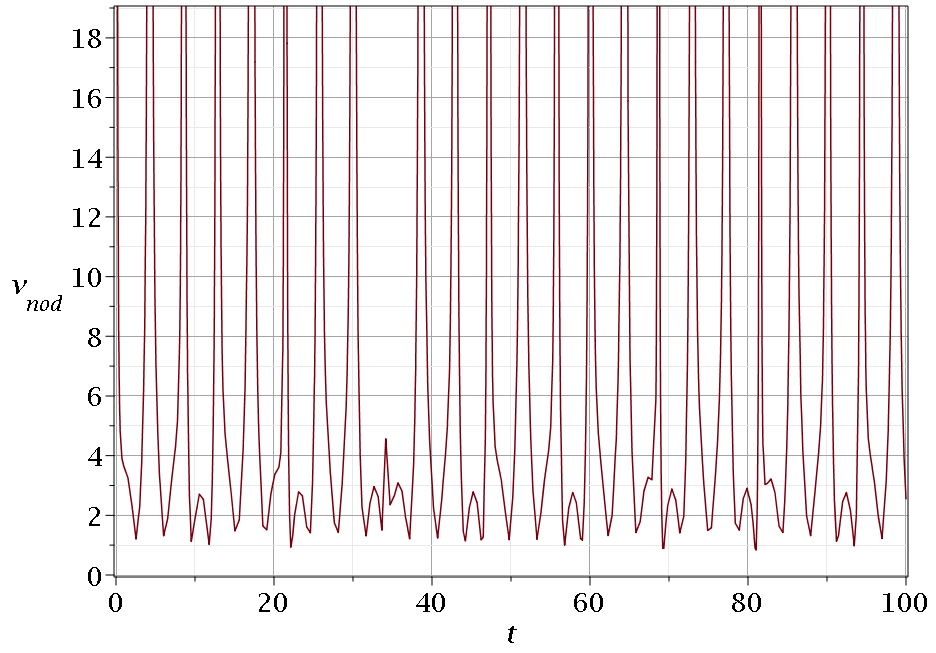}
\caption{The absolute value of the velocity of a certain nodal point of the state $\Psi$ with 
$k=1, \omega_x=1,\omega_y=\sqrt{3}$. We 
observe the repeated fast motion to infinity.}
\label{vnodal}
\end{figure}

\begin{figure}[h]
\centering
\includegraphics[scale=0.23]{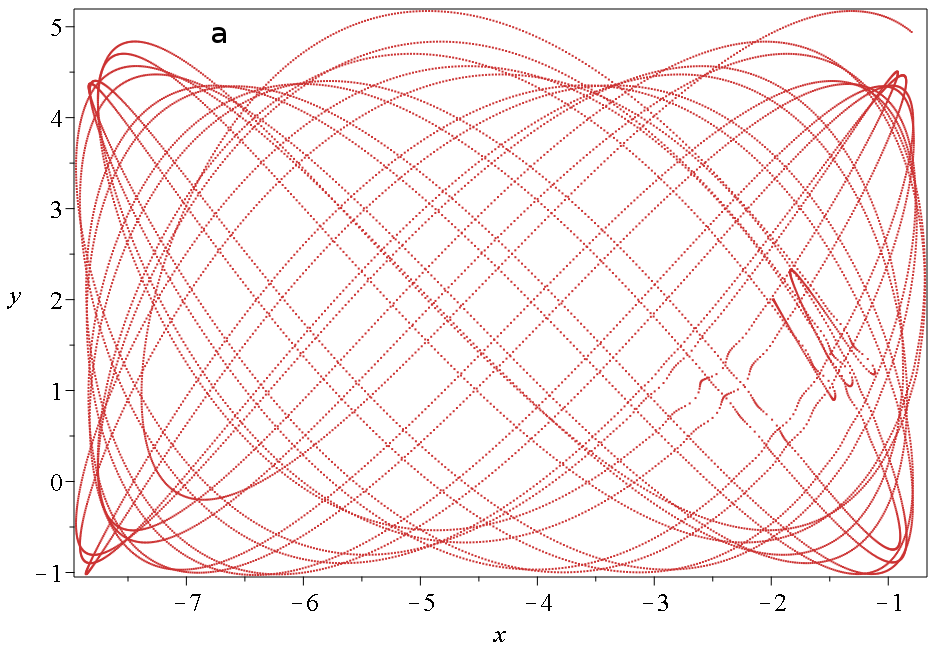}
\includegraphics[scale=0.23]{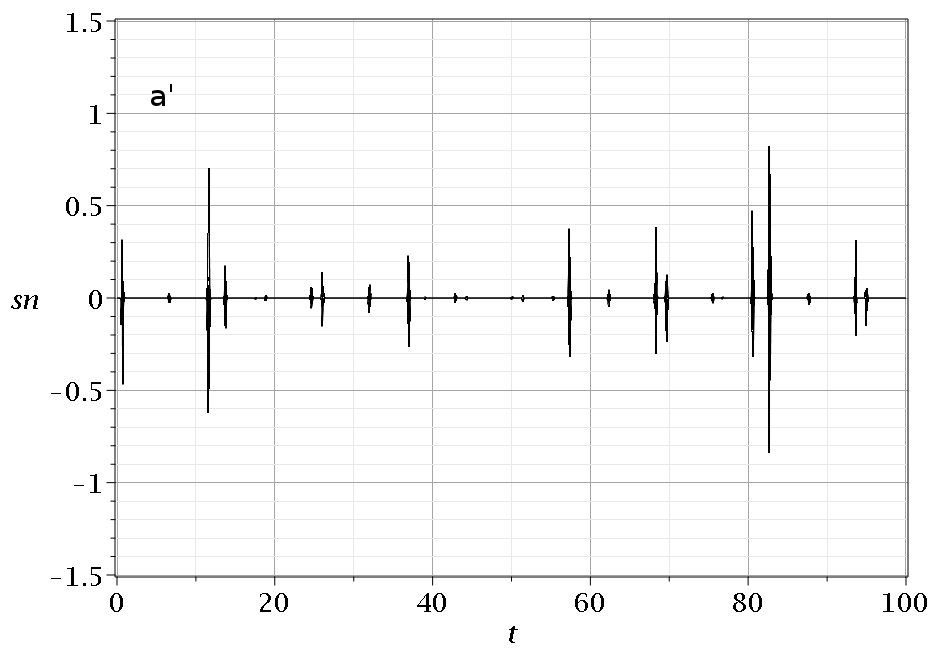}
\includegraphics[scale=0.23]{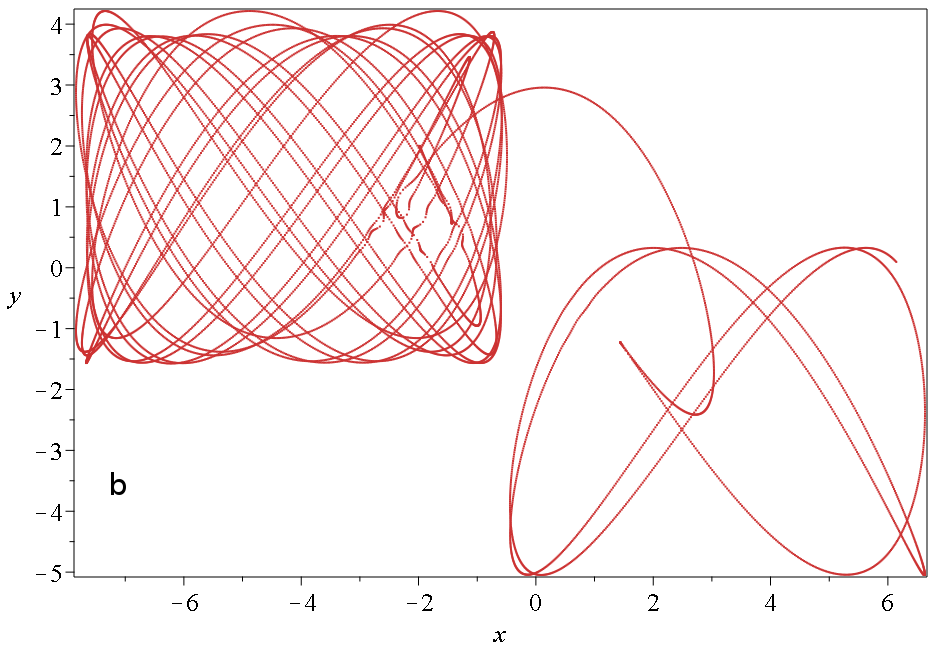}
\includegraphics[scale=0.23]{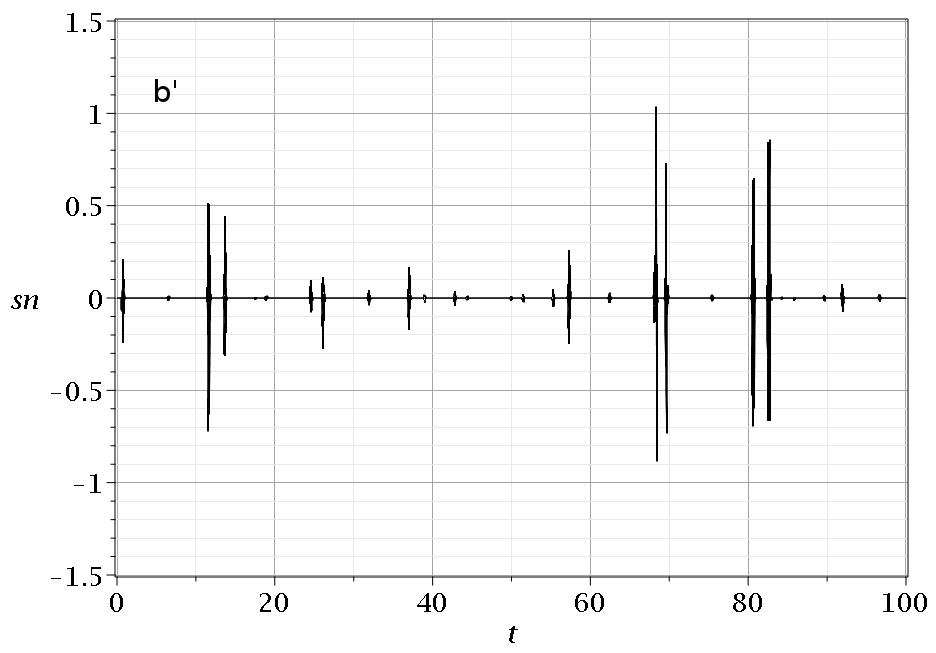}
\includegraphics[scale=0.23]{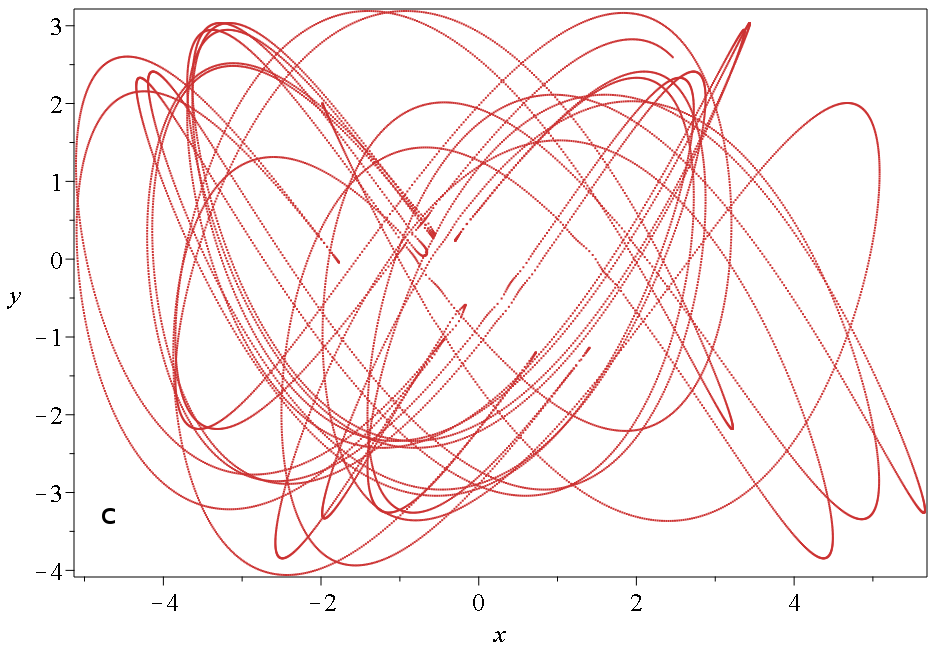}
\includegraphics[scale=0.23]{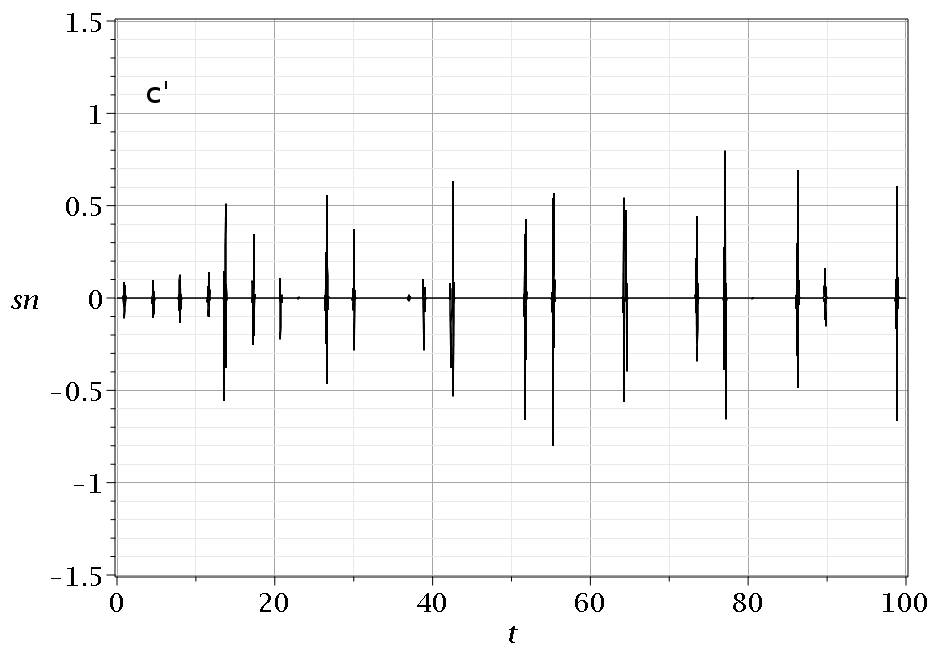}
\caption{Bohmian trajectories and the corresponding variations of the stretching numbers for the state $\Psi$. Entanglement brings chaos and changes significantly
the behaviour of the Bohmian trajectory from that of a Lissajous
curve. 
a) If we set $c_2=2\times 10^{-6}$ we see that 
the initial Lissajous  curve becomes deformed 
and its dimensions change. When the orbit approaches an X-point we have an abrupt change of the stretching number. b)  For $c_2=2\times 10^{-5}$ the behaviour 
changes drastically. At time $t=82.66$ there is a strong 
scattering event (b') which derails the trajectory outside 
the initial region.
c) If we increase entanglement chaos becomes apparent
 immediately as in the case of a maximally 
 entangled state with $c_2=\sqrt{2}/2$. $(x(0)=-2, 
y(0)=2, \omega_x=1, \omega_y=\sqrt{3})$, which is followed by many abrupt changes of the stretching number (c'). }
\label{ektropi}
\end{figure}

This behaviour changes with the onset of 
entanglement, namely with the increase of $c_2$. 
In Fig.~\ref{diaforetika_k} we show the 
trajectories of two different nodal points 
with $k=1$ and $k=3$ for the wavefunction $\Psi$. The  frequencies are 
$\omega_x=1$ and $\omega_y=\sqrt{3}$. The coefficient
$c_2$ is set equal to $2\times 10^{-5}$, 
namely the entanglement
is extremely small, $EE\simeq 8\times 10^{-10}$.
Even with this slight perturbation we 
observe that the trajectories of the 
nodal points enter repeatedly the region of 
space where the 
support of the wavefunction is strong and 
then go repeatedly to infinity with 
very high velocities (Fig.~\ref{vnodal}).
Whenever a Bohmian 
particle comes close to one of
the infinitely many NPXPCs, there is a sudden spike in the evolution
of the stretching number.
Consequently, chaos emerges
as shown in Fig.~\ref{ektropi}.

Figure \ref{ektropi}a,a' shows what 
happens in the case of the 
state $\Psi$ when $c_2=2\times10^{-6}$: 
the initial Lissajous curve becomes deformed 
and its size changes. However 
the trajectory still evolves in a almost 
confined region of physical space up for
$t\in [0,100]$ and the action of the NPXPCs 
is limited to the deformation of the 
initial Lissajous 
curve as we can see in the scattering 
events for $-2\leq x\leq -1 $ and $1\leq y\leq 2$, 
which correspond to the initial 
shifts of the stretching number. 

\begin{figure}[h]
\centering
\includegraphics[scale=0.35]{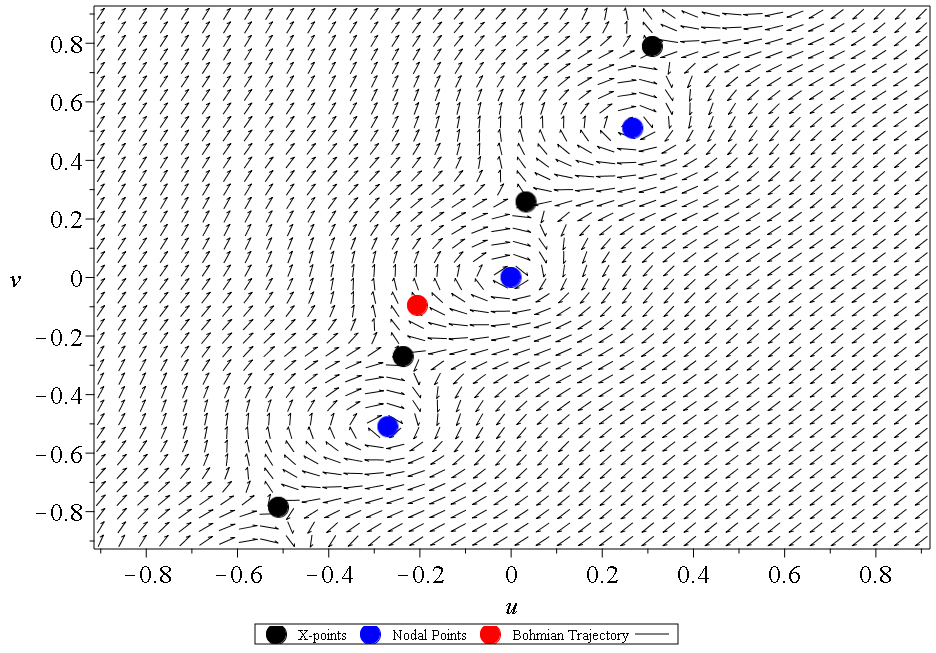}
\caption{The NPXPCs of the state $\Psi$ at $t=t_d\simeq 82.66 $. 
We observe that the Bohmian particle (red point) 
is close to one of the X-points. The interaction
of this particle with the X-point
produces a jump in the stretching number. 
Such close encounters between particles and 
X-points lead to a positive LCN, and consequently
to chaos $(x(0)=-2, 
y(0)=2, \omega_x=1, \omega_y=\sqrt{3},c_2=2\times 10^{-5})$}
\label{event}
\end{figure}

However for $c_2=2\times10^{-5}$ 
(Fig.~\ref{ektropi}b,b') besides 
the deformation of the initial
Lissajous type curve, there is a strong 
scattering event which forces the 
trajectory to exit the initial region and 
move to an other region on the lower right
part of the figure with almost the same 
horizontal and vertical dimensions. We call `derailment time' the time when such a major scattering event takes place. In Fig.~\ref{ektropi} the
derailment time is $t_d\simeq 82.66$. 
Indeed the stretching number undergoes
a strong positive shift at $t_d$. 
This  is shown 
in detail in Fig.~\ref{event}. 
Multiple NPXPCs cross the plane 
$(x,y)$ at $t_d$. The 
nodal points are colored blue, the
X-points are black. The Bohmian trajectory is close to 
one of the X-points of the multiple
NPXPCs, hence subject to scattering.

\begin{figure}[h]
\centering
\includegraphics[scale=0.35]{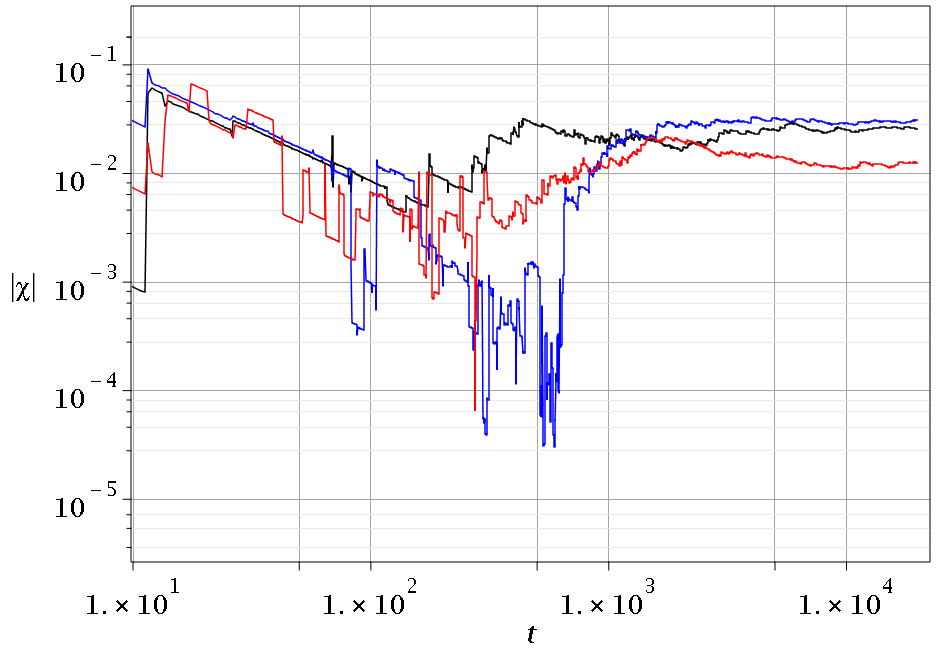}
\caption{The absolute value of the  finite time Lyapunov characteristic number $\chi$ for the three cases of Fig.~\ref{ektropi} (state $\Psi$) in logarithmic axes. The blue curve corresponds to $c_2=2\times 10^{-6}$, the black curve corresponds to $c_2=2\times 10^{-5}$ and the red curve corresponds to $c_2=\sqrt{2}/{2}$. The sudden droppings of $|\chi(t)|$  to very small values (on  a logarithmic scale) correspond to the crossings of the axis $\chi(t)=0$ during the oscillatory motion of $\chi(t)$ between positive and negative values. However, we observe that for  times larger than $t=1.5\times 10^4$  $\chi(t)$ stabilizes at a positive value. Consequently the trajectories are chaotic. }
\label{FLCN}
\end{figure}

Increasing $c_2$ in the extreme case of a maximally entangled 
state $(c_1=c_2=\sqrt{2}/2)$ we observe chaos 
immediately through a large number of 
scattering events, as shown in Fig.~\ref{ektropi}c,c'
and the corresponding  time series of the 
stretching number
\cite{Contopoulos200210}).
The finite time Lyapunov characteristic number $\chi$ for all cases is given
in Fig.~\ref{FLCN}.

\subsection{Commensurable frequencies and the special case of isotropic oscillators}

In Fig.\ref{basis_states} we show the oscillatory
behavior of the absolute value of two
one-dimensional coherent states with the 
same frequency $\omega_x=\omega_y=1$
and amplitude 
$|A_0|=5/2$ which at 
$t=0$ start from 
the right and the left limiting points of 
the oscillation (blue and red curve respectively).
This is done by choosing $\sigma=0$ for the
first one and $\sigma=\pi$ for the second one.
We observe that at $t=\pi/3$  the two curves 
have a complete spatial overlap at $x=0$ 
(black curve), which is totally different 
from their almost vanishing overlap in phase
space.

\begin{figure}[h]
\centering
\includegraphics[scale=0.23]
{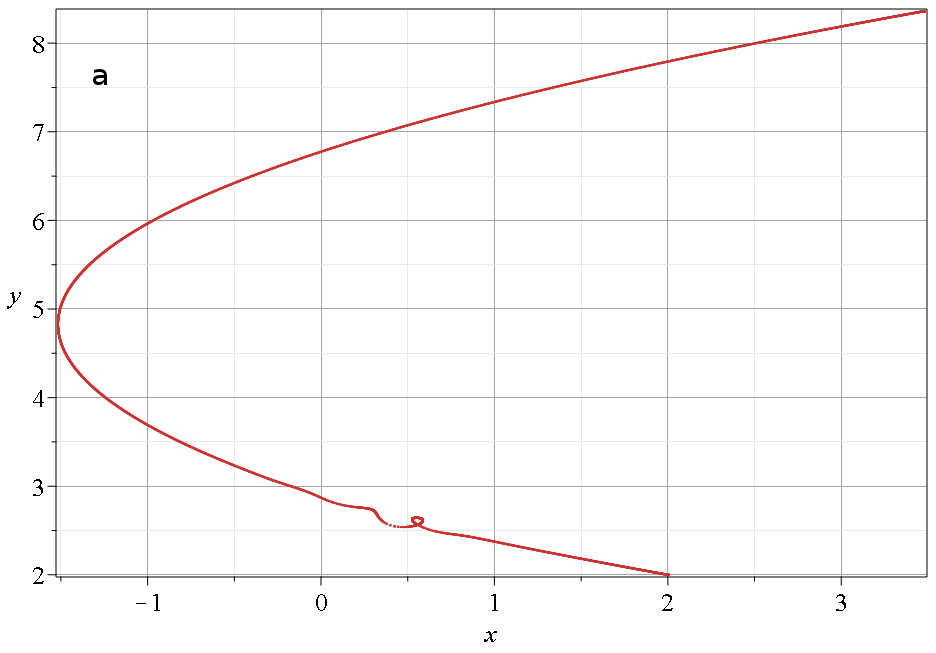}
\includegraphics[scale=0.23]{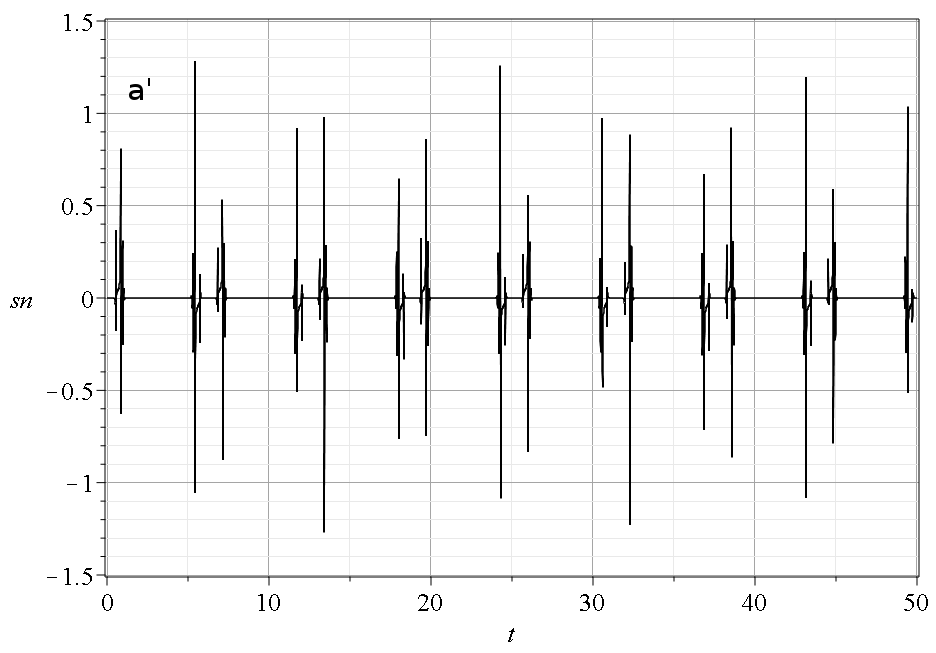}
\includegraphics[scale=0.23]{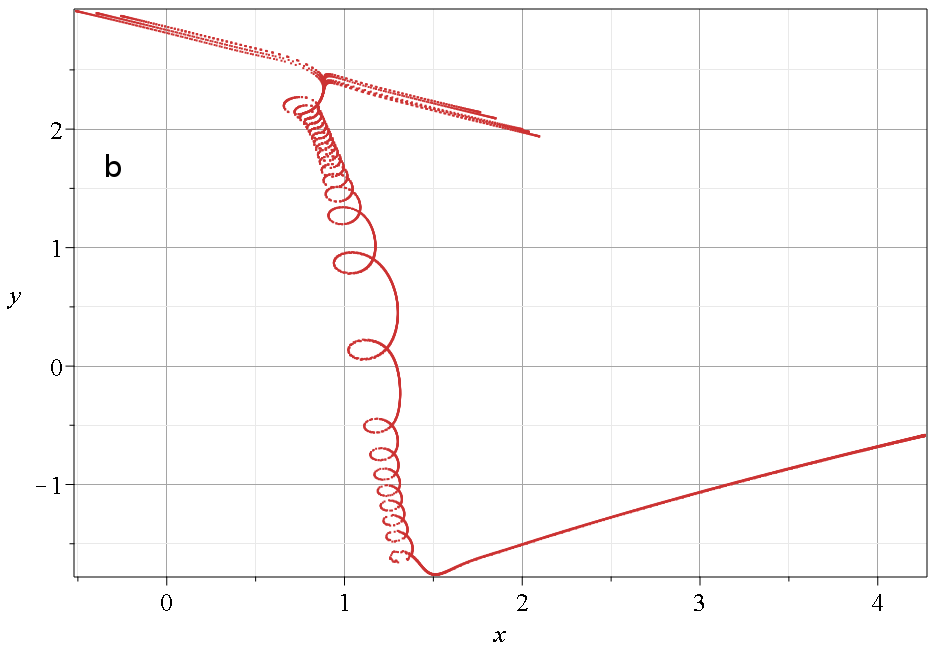}
\includegraphics[scale=0.23]{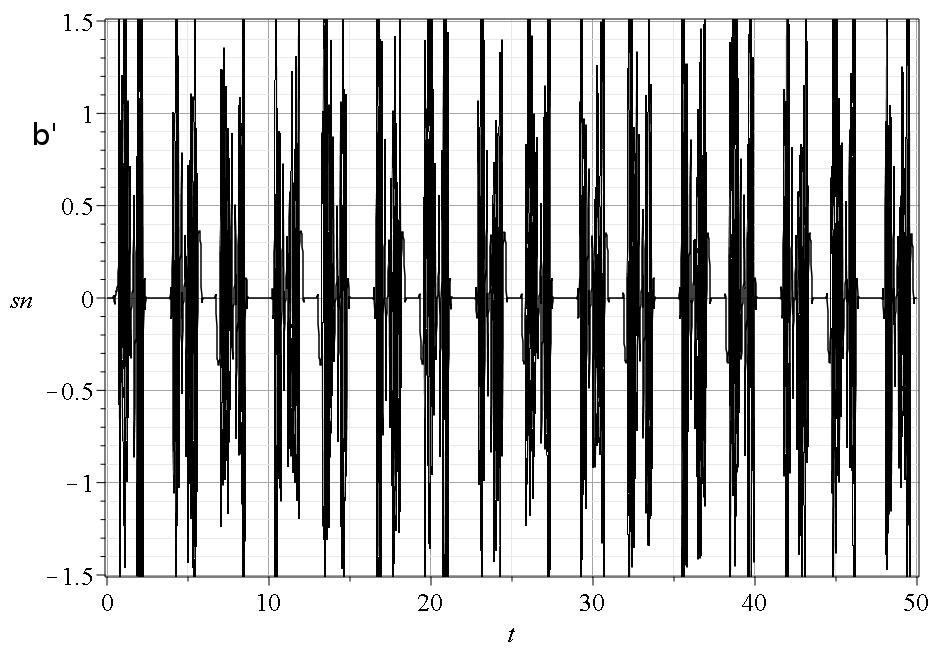}
\includegraphics[scale=0.23]
{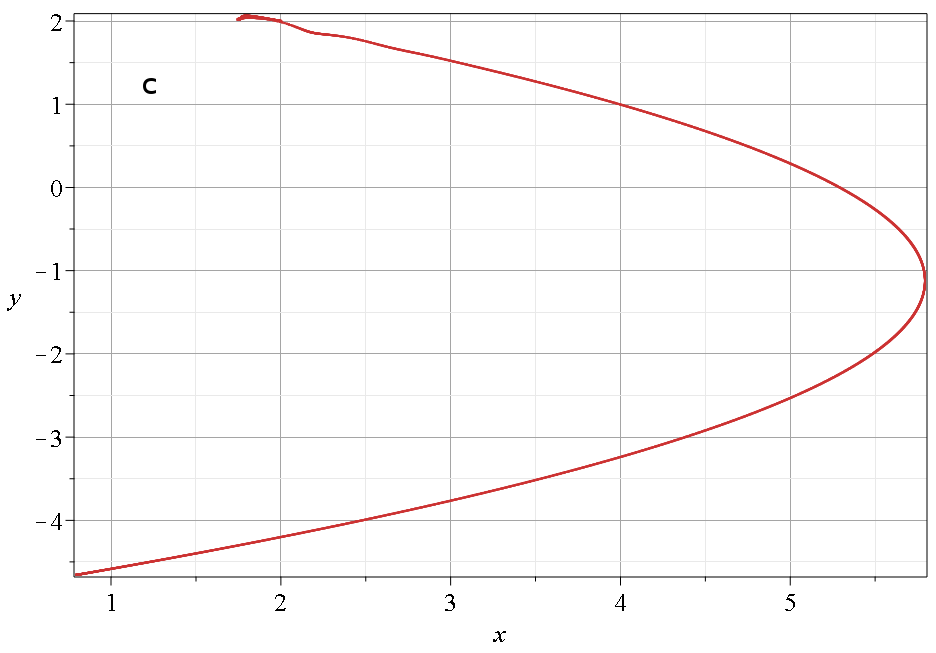}
\includegraphics[scale=0.23]{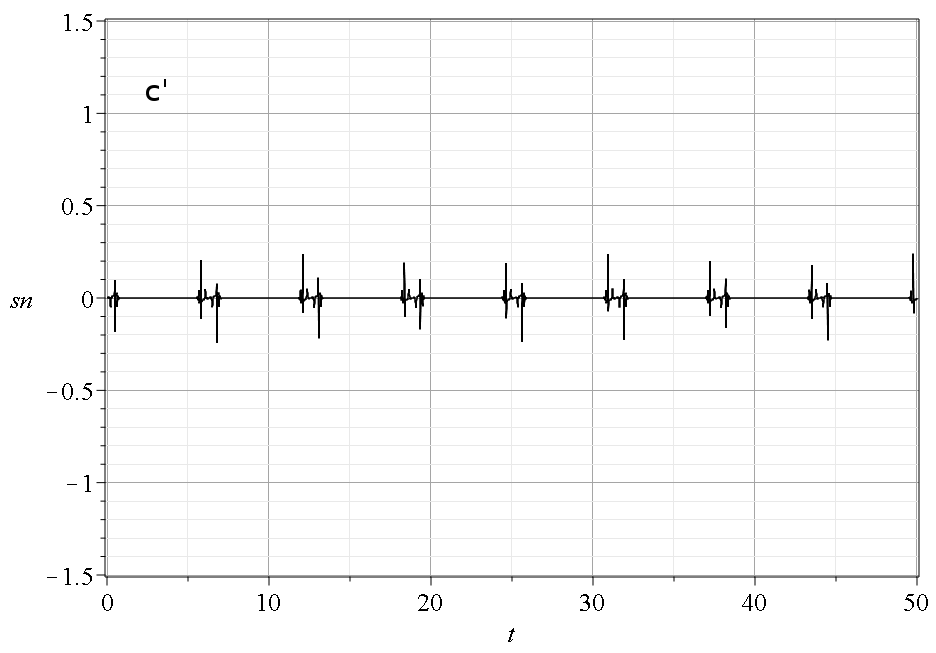}
\caption{Three Bohmian trajectories of the state $\Psi$ with common initial 
condition $x(0)=y(0)=2$ for $\omega_x=2, \omega_y=1$. 
a) $c_2=2\times 10^{-6}$, b) $c_2=2\times 10^{-5}$ and c) $c_1=c_2=\sqrt{2}/2$. The evolution of the corresponding stretching numbers is shown in (a'), (b') and (c') respectively. The
spikes in the latter plots correspond to scattering events of the trajectories by the NPXPCs. These trajectories are all periodic, but they exhibit considerable complexity due to the interaction with the NPXPCs. }
\label{ektropi_rita}
\end{figure}

In Fig.~\ref{ektropi_rita} we present a case where
$\omega_x=2, \omega_y=1$ 
for a certain initial condition $x(0)=y(0)=2$ and 
for different values of $c_2$ in the state $\Psi$. We note that in 
the absence of entanglement, namely in the case 
of a product state, the trajectory is periodic, 
since it is the composition of two oscillations 
with $\omega_x/\omega_y$ is a rational number.
The insertion of entanglement implies the
existence of nodes. However, with some
algebra one can see that Eqs.~(\ref{rational1}) and (\ref{rational2})
yield periodic solutions. In fact if $\omega_1/\omega_2$ is an irreducible ratio $s_1/s_2$, where $s_1, s_2$ are  positive integers, then the period of the system is $T=2\pi s_2/\omega_y$. Moreover there
is time reversal antisymmetry, namely
for $t'= -t$ we get $dx/dt'=-dx/dt$ and $dy/dt'=-dy/dt$.  Finally for $t=0$ we have $dx/dt=dy/dt=0$. Consequently the motion is periodic with reflection at $t=0$. The same property holds for 
the  state $\Phi$.

In Fig.~\ref{ektropi_rita}a,a', 
while the amount of entanglement is extremely 
small, one observes strong scattering events in 
the stretching numbers. For a larger,  
but still small, value of 
entanglement, we observe 
spiral motion (Fig.~\ref{ektropi_rita}b,b'). This is the typical behaviour of 
a Bohmian trajectory close to a moving nodal point. 
In this case scattering effects are very strong. 
Finally for a maximally entangled state the scattering effects are milder (Fig.~\ref{ektropi_rita}c,c'). 

\begin{figure}[h]
\centering
\includegraphics[scale=0.3]{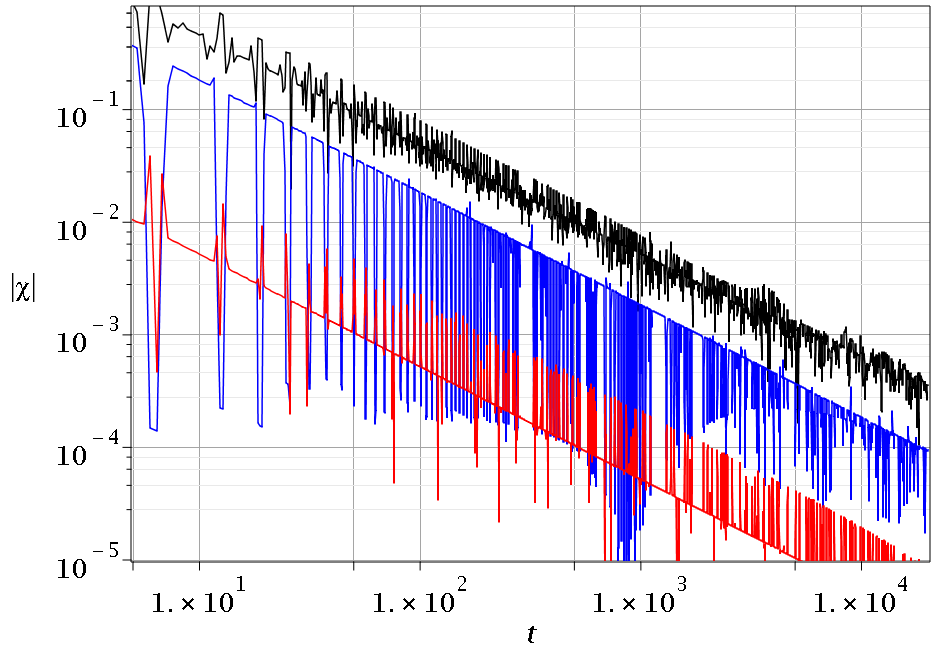}
\caption{The absolute value of the finite time Lyapunov characteristic number $\chi$ for the three cases of Fig.~\ref{ektropi_rita} (state $\Psi$) in logarithmic axes. The blue curve corresponds to $c_2=2\times 10^{-6}$, the black curve corresponds to $c_2=2\times 10^{-5}$ and the red curve corresponds to $c_2=\sqrt{2}/{2}$. We observe the decrease of $\chi$ over the course of time according to a power law, which is a signature of ordered trajectories.  Although there exist NPXPCs there is no chaos production, since the motion is periodic.}
\label{FLCN2}
\end{figure}

In all cases $\chi$ vanishes in the course of
time. In Fig.~\ref{FLCN2}, the blue curve corresponds to $c_2=2\times 10^{-6}$, the  black curve corresponds to $c_2=2\times 10^{-5}$ and the red curve corresponds to  $c_2=\sqrt{2}/{2}$. 
However, the system may appear as `effectively chaotic' ($\chi$ large) for transient times long enough but shorter than the period. Conversely, a system with incommensurable ratio $\omega_x/\omega_y$ may appear as `effectively ordered' ($\chi$ going temporarily to zero) at times corresponding to approximate periods defined by rational approximation of the ratio $\omega_x/\omega_y$.  For the details about the distinction between effectively chaotic and effectively ordered  orbits see \cite{contopoulos2008ordered}.

In the extreme case where $\omega_x=\omega_y$,
namely when the oscillators are isotropic,
we get for the state $y_{nod}/x_{nod}=1$ for the state $\Psi$. 
From the common denominator of Eqs.
(\ref{xnod}) and  (\ref{ynod}), 
 both $x$ and $y$ go to infinity
when $\omega_{xy}=0$.  Consequently as 
$\omega_{xy}\to 0$
the nodal points disappear very fast 
along the line $y=x$ and we find ordered trajectories. 
Similarly for the state $\Phi$ we 
have $(x,y)\to\infty$ as $\omega_x\to\omega_y$ 
with $y_{nod}/x_{nod}=-1$ and again we 
find ordered trajectories.

\begin{figure}
\centering
\includegraphics[scale=0.35]{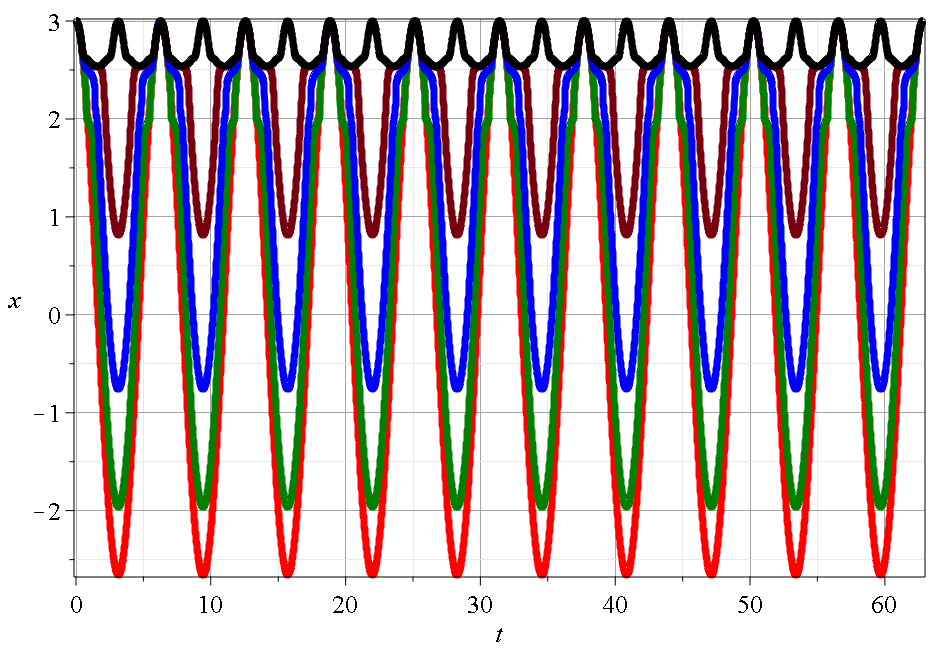}
\caption{The motion of the Bohmian particle in the case $\omega_x=\omega_y=1$ for a given initial condition and different amounts of entanglement (state $\Psi$). Red curve: $c_2=2\times 10^{-5}$, green curve: $c_2=2\times 10^{-2}$, blue curve: $c_2=4\times 10^{-1}$ burgundy curve: $c_2=0.701$ and black curve: $c_2=\sqrt{2}/2\simeq 0.707$. The larger the entanglement, the smaller the range of motion. The period of all orbits is $2\pi$ except from the maximal entanglement case (black curve), where is equal to $\pi$. In this case, the entanglement introduces no chaos but affects the Fourier spectrum of the periodic trajectories.}
\label{talantoseis}
\end{figure}

In fact one can easily 
check that in the case of the isotropic 
oscillators both states produce ordered
trajectories characterized 
by integrals of motion: the state 
$\Phi$ gives $y-y(0)=x-x(0)$ 
while $\Psi$ gives $y-y(0)=-(x-x(0))$. 
Consequently the particles move on 
certain straight lines parallel to the 
diagonals $x=\pm y$ defined from their initial
positions. Here the effect of the increase
of the entanglement is a gradual deformation 
and spatial 
confinement of the oscillatory motion
which occurs when the system is separable ($c_2=0$). 
In Fig.~\ref{talantoseis} we observe 
this deformation for the same initial 
value $x(0)=3, y(0)=2$ and different values of $c_2$ in the state $\Psi$.
For small values of $c_2$ (small entanglement) the trajectory
is similar to that of the separable case, with 
only a mild deformation at $x\simeq2$. 
 The range of motion in the x-axis
gets smaller with the increase of $c_2$, while the range of motion 
in the y-axis is affected via the integral of motion
$y=y(0)-(x-x(0))$ and becomes smaller with the increase of entanglement as well. Moreover we observe that in the partially 
entangled states the leading Fourier component is $\sin(t)$, while higher harmonics ($\sin(mt),\, m=2,3,\dots)$ become increasingly important as the entanglement increases. In fact, the leading Fourier term
in the maximally entangled state is $\sin(2t)$.

\begin{figure}
\centering
\includegraphics[scale=0.35]{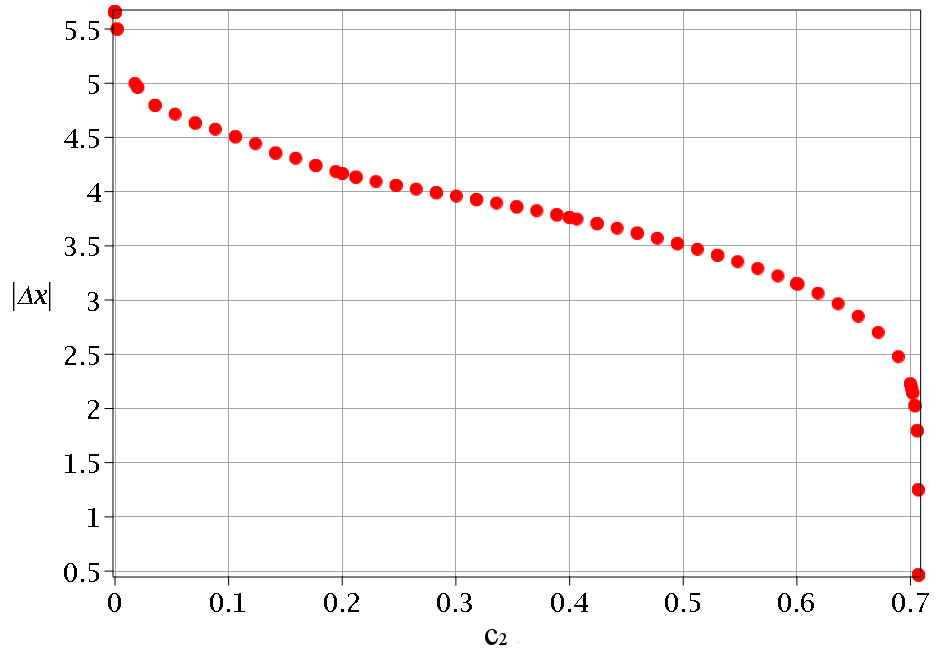}
\caption{The range of motion $|\Delta x|$ in the x-axis as a function 
of the coefficient $c_2$ (and consequently of entanglement) in the case of Fig.~\ref{talantoseis}.
We observe the fast decrease of $|\Delta x|$ in the regions of small and large values
of $c_2$. For intermediate values of $c_2$ the range of motion exhibits an approximatively linear decrease. Entanglement differentiates clearly the trajectories in the extreme cases of $c_2=0$ (product state) where  $|\Delta x|=5.657$ and  for $c_2=\sqrt{2}/2$ (maximally entangled state) where $\Delta x$ is minimum and equal to $0.464$. }
\label{range}
\end{figure}

We note here that
the integration of these seemingly simple ordered trajectories becomes
difficult when we approach the maximally entangled state where 
$c_1=c_2=\sqrt{2}/2$, due to the high degree of nonlinearity.
For $c_2\in[0.7,\sqrt{2}/2\simeq 0.707106781]$ one needs
very accurate numerical integration since even with a small 
time step, for example a 4th order Runge-Kutta scheme with
time step of order $\mathcal{O}(10^{-4})$, there is a large 
accumulative error over the course of time, which leads to 
erroneous non-periodic solutions with respect to the initial
conditions, namely to trajectories that do not reach the initial
conditions in every cycle. In  the case 
of adaptive numerical schemes, like 4th order Runge-Kutta-Fehlberg with 
5 order error estimator, one needs very small tolerance in absolute 
and relative errors, even for short time monitoring of the solution.
In order to monitor the trajectory correctly in the relatively
wide time window
of the first 10 cycles of motion ($0\to 20\pi$), we applied the 
Dormand Prince method  of
8th order with 5th order error estimator, and with absolute error
tolerance $atol=10^{-18}$ and relative error tolerance
$rtol=10^{-17}$ (\cite{Dormand}). This peculiar behaviour of the system
in the close region of maximally entangled state stems from the high contribution of all nonlinear terms of the equations of motion, which in the case of maximally entangled state
becomes of the same order of magnitude.  
In Fig.~\ref{range} we present the variation of $|\Delta x|$ for 
different values of the coefficient $c_2$. In the cases 
of small and large entanglement we observe a fast decrease
of $|\Delta x|$, while for intermediate values of $c_2$
the decrease is almost linear. 

 To sum up, we see that in the 
presence of integrability the QE affects the trajectories
differently than in the case of non-integrability.
It does not  change significantly the shape of the joint evolution, but
it dictates the range of motion in both axes.

\section{Conclusions}\label{conc}

In the present work we studied the effect of quantum entanglement (QE) on 
the Bohmian trajectories of a simple bipartite 
system, composed of two entangled one dimensional
canonical coherent states.  Our main results are:
\begin{enumerate}
\item{The presence of entanglement plays a crucial 
role for the evolution of Bohmian particles. In 
fact a very small amount of entanglement which 
can be thought of as a small perturbation in a 
product state, suffices to bring strong chaos by 
activating the nodal point-X-point mechanism.}
\item{A simple (but highly non-linear) 
two-qubit Bohmian system has infinitely many 
nodal point-X-point complexes (NPXPCs) which not only
cover large parts of physical space, but also
evolve very fast in time and scatter most of the
Bohmian trajectories. Consequently for  
every non-vanishing value of entanglement the system is characterized by complex dynamics. }
\item{They key parameter of the system is the frequency ratio $\omega_1/\omega_2$. In the case of anisotropic oscillators with incommensurable frequencies entanglement is a prerequisite for the existence of NPXPCs which scatter the trajectories and produce chaos. }
\item{In the case of anisotropic oscillators with commensurable frequencies we found that the trajectories are complicated but 
periodic. Consequently we have `effectively chaotic' trajectories for
transient times smaller than the period $T$ of the system, but the motion turns
eventually to be ordered.}
\item When the oscillators are isotropic
the NPXPCs dissapear (go to infinity) and the 
system is integrable. The trajectories are
confined to certain straight lines depending on
the initial conditions. In that case the presence
of entanglement can not  affect 
the integrability of the system, but it  
changes the shape of the trajectories. We monitored the range of motion as a function of the coefficient $c_2$ and found that the larger the $c_2$, the smaller the $|\Delta x|$. For small and large values of $c_2$ the variation of $|\Delta x|$
was fast, while for intermediate values it was almost linear with respect to $c_2$. Furthermore, the increase of the entanglement implies a change in the Fourier components of the periodic motion.
In the extreme case of a maximal entangled state the leading Fourier 
term changes from $\sin(t)$ to $\sin(2t)$.
\end{enumerate}

In this paper we connected the  evolution 
of Bohmian trajectories with the degree
of the entanglement of their guiding 
wavefunction. We worked with the simplest non-trivial case, namely  two entangled qubits. According to our remarks (iii) and (iv) above, the  entanglement leads in most cases to chaos, in agreement with the
results of \cite{cesa2016chaotic}. However, there are also cases in which the entanglement does not produce chaos, but it can still affect the spectrum of the regular trajectories. 
 An interesting question for further study is the relation between the chaotic/regular Bohmian trajectories and the (possibly) conserved mean values of several quantities such as energy, linear momentum, angular momentum etc.\cite{holland1995quantum} in entangled bipartite systems.

The simple wavefunctions of the present paper provide useful information about the phenomenology
of the entanglement in Bohmian trajectories. This is  a necessary step towards
the  exploitation of Bohmian trajectories
for a  trajectory-based
characterization of QE (construction of indicators
and measures), with possible applicability in the case of high-dimensional 
bipartite systems, or multipartite systems, where the quantification
of entanglement remains an open problem.

\clearpage
\ack{This research is supported by the 
Research Commitee of the Academy of Athens.}

\section*{References}
\bibliographystyle{iopart-num}
\bibliography{bibliography}

\providecommand{\newblock}{}
\begin{thebibliography}{10}
\expandafter\ifx\csname url\endcsname\relax
  \def\url#1{{\tt #1}}\fi
\expandafter\ifx\csname urlprefix\endcsname\relax\def\urlprefix{URL }\fi
\providecommand{\eprint}[2][]{\url{#2}}

\bibitem{Bohm}
Bohm D 1952 {\em Phys. Rev.\/} {\bf 85}(2) 166

\bibitem{BohmII}
Bohm D 1952 {\em Phys. Rev.\/} {\bf 85}(2) 180

\bibitem{horodecki2009quantum}
Horodecki R, Horodecki P, Horodecki M and Horodecki K 2009 {\em Rev. Mod.
  Phys.\/} {\bf 81} 865

\bibitem{mintert2005measures}
Mintert F, Carvalho A~R, Ku{\'s} M and Buchleitner A 2005 {\em Phys. Rep.\/}
  {\bf 415} 207--259

\bibitem{nielsen2004quantum}
Nielsen M~A and Chuang I~L 2004 {\em Quantum Computation and Quantum
  Information (Cambridge Series on Information and the Naturciences)\/}
  (Cambridge University Press)

\bibitem{durt2002bohm}
Durt T and Pierseaux Y 2002 {\em Phys. Rev. A\/} {\bf 66} 052109

\bibitem{braverman2013proposal}
Braverman B and Simon C 2013 {\em Phys. Rev. Lett.\/} {\bf 110} 060406

\bibitem{norsen2014weak}
Norsen T and Struyve W 2014 {\em Ann. Phys.\/} {\bf 350} 166--178

\bibitem{mahler2016experimental}
Mahler D~H, Rozema L, Fisher K, Vermeyden L, Resch K~J, Wiseman H~M and
  Steinberg A 2016 {\em Science advances\/} {\bf 2} e1501466

\bibitem{elsayed2018entangled}
Elsayed T~A, M{\o}lmer K and Madsen L~B 2018 {\em Scient. Rep.\/} {\bf 8} 12704

\bibitem{cesa2016chaotic}
Cesa A, Martin J and Struyve W 2016 {\em J. Phys. A\/} {\bf 49} 395301

\bibitem{de2012bohmian}
de~Almeida A, de~Ponte M, Cardoso W, Avelar A, Moussa M and de~Almeida N 2012
  {\em arXiv preprint arXiv:1204.6314\/}

\bibitem{ramvsak2012spin}
Ram{\v{s}}ak A 2012 {\em J. Phys. A\/} {\bf 45} 115310

\bibitem{garrison2008quantum}
Garrison J and Chiao R 2008 {\em Quantum optics\/} (Oxford University Press)

\bibitem{makarov2018coupled}
Makarov D~N 2018 {\em Phys. Rev. E\/} {\bf 97} 042203

\bibitem{zander2018revisiting}
Zander C and Plastino A 2018 {\em Entropy\/} {\bf 20} 473

\bibitem{voglis1994invariant}
Voglis N and Contopoulos G 1994 {\em J. Phys. A\/} {\bf 27} 4899

\bibitem{Contopoulos200210}
Contopoulos G 2002 {\em Order and Chaos in Dynamical Astronomy\/} (Springer)

\bibitem{efthymiopoulos2006chaos}
Efthymiopoulos C and Contopoulos G 2006 {\em J. Phys. A\/} {\bf 39} 1819

\bibitem{PhysRevE.79.036203}
Efthymiopoulos C, Kalapotharakos C and Contopoulos G 2009 {\em Phys. Rev. E\/}
  {\bf 79} 036203

\bibitem{tzemos2018integrals}
Tzemos A~C and Contopoulos G 2018 {\em J. Phys. A\/} {\bf 51} 075101

\bibitem{tzemos2018origin}
Tzemos A~C, Efthymiopoulos C and Contopoulos G 2018 {\em Phys. Rev. E\/} {\bf
  97} 042201

\bibitem{Wisniacki1}
Wisniacki D~A and Pujals E~R 2005 {\em Europhys. Lett.\/} {\bf 71} 159

\bibitem{Wisniacki2}
Wisniacki D~A, Pujals E~R and Borondo F 2007 {\em J. Phys. A\/} {\bf 40} 14353

\bibitem{contopoulos2008ordered}
Contopoulos G and Efthymiopoulos C 2008 {\em Celest. Mech. Dyn. Astron.\/} {\bf
  102} 219

\bibitem{Dormand}
Dormand J~R 1996 {\em Numerical Methods for Differential Equations: A
  Computational Approach\/} (CRC Press)

\bibitem{holland1995quantum}
Holland P~R 1995 {\em The quantum theory of motion: an account of the de
  Broglie-Bohm causal interpretation of quantum mechanics\/} (Cambridge
  University Press)

\end{thebibliography}

\end{document}